\documentclass[aps,prb,twocolumn,superscriptaddress,nobibnotes,longbibliography]{revtex4-2}
\bibpunct{[}{]}{;}{n}{}{}

\usepackage{graphics, bm, xspace}
\usepackage{graphicx}
\usepackage{psfrag}
\usepackage{amsmath}
\usepackage{amssymb}
\usepackage{epsfig}
\usepackage{subfigure}
\usepackage{grffile}
\usepackage{times}
\usepackage{color}
\usepackage{multirow}
\usepackage[bookmarks=false,linkcolor=blue,urlcolor=blue,colorlinks,citecolor=blue]{hyperref}
\usepackage{float}
\usepackage{gensymb}
\usepackage{mhchem} 

\newcommand{\beq}    {\begin{equation}}
\newcommand{\enq}    {\end{equation}}
\newcommand{\ceq}[1] {(\ref{#1})}

\newcommand{\eps}{\epsilon}
\newcommand{\kk}{{\bf k}}

\newcommand{\rr}{{\bf r}}

\newcommand{\qq}{{\bf q}}

\newcommand{\KK}{{\bf K}\xspace}

\newcommand{\GGv}{{\bf G}\xspace}

\newcommand{\df}     {\equiv}

\newcommand{\GG}     {$Q_{\mu\nu}$\xspace}
\newcommand{\ReGG}   {${\rm Re}[Q_{\mu\nu}]$\xspace}
\newcommand{\dsm}    {D^{\rm(s)}_{\mu\nu}}
\newcommand{\ds}     {$\dsm$\xspace}

\newcommand{\gmm}    {g_{\mu\nu}}
\newcommand{\gm}     {$\gmm$\xspace}

\newcommand{\xx}     {{\bf x}}
\newcommand{\aaa}    {{\bf A}\xspace}
\newcommand{\ket}[1] {\lvert #1 \rangle}
\newcommand{\bra}[1] {\langle #1 \rvert}

\newcommand{\braket}[2]  {\langle #1 \rvert #2 \rangle}
\newcommand{\tkt}     {$T_{KT}$\xspace}
\newcommand{\rhosm}    {\rho^{\rm (s)}_{\mu\nu}}
\newcommand{\rhos}     {$\rho^{\rm (s)}_{\mu\nu}$\xspace}

\newcommand{\rhosam}   {\rho^{\rm (s,1)}_{\mu\nu}}
\newcommand{\rhosa}    {$\rho^{\rm (s,1)}_{\mu\nu}$\xspace}
\newcommand{\rhosbm}   {\rho^{\rm (s,2)}_{\mu\nu}}
\newcommand{\rhosb}    {$\rho^{\rm (s,2)}_{\mu\nu}$\xspace}
\newcommand{\rhoscm}   {\rho^{\rm (s,3)}_{\mu\nu}}
\newcommand{\rhosc}    {$\rho^{\rm (s,3)}_{\mu\nu}$\xspace}

\newcommand{\rhosg}    {$\rho^{\rm (s,geo)}_{\mu\nu}$\xspace}
\newcommand{\rhoslg}   {$\rho^{\rm (s,geo)}_{\mu\mu}$\xspace}
\newcommand{\rhosconv} {$\rho^{\rm (s,conv)}_{\mu\nu}$\xspace}
\newcommand{\rhoslconv} {$\rho^{\rm (s,conv)}_{\mu\mu}$\xspace}
\newcommand{\rhosspin}{$\rho^{\rm(s,spin)}_{\mu\nu}$\xspace}
\newcommand{\hh}{$\cal{H}$\xspace}
\newcommand{\ph}{$\cal{P_H}$\xspace}
\newcommand{\hbdg}{H_{\rm BdG}}
\newcommand{\emt}{\epsilon_{m_T}^{(T)}}
\newcommand{\ent}{\epsilon_{n_T}^{(T)}}
\newcommand{\mt}{\lvert m_T \rangle}

\newcommand{\emb}{\epsilon_{m_B}^{(B)}}

\newcommand{\epb}{\epsilon_{p_B}^{(B)}}
\newcommand{\eqb}{\epsilon_{q_B}^{(B)}}
\newcommand{\emx}{\epsilon_{m_X}^{(X)}}
\newcommand{\enx}{\epsilon_{n_X}^{(X)}}
\newcommand{\mb}{\lvert m_B \rangle}

\newcommand{\hs}{\cal{H}}
\newcommand{\tr}{{\rm Tr}}

\begin{document}


\title{Quantum Metric and Correlated States in Two-dimensional Systems}
\author{Enrico Rossi}
\email[Corresponding author: ]{erossi@wm.edu}
\affiliation{Department of Physics, William \& Mary, Williamsburg, Virginia (23187), USA}

\begin{abstract}
The recent realization of twisted, two-dimensional, bilayers exhibiting strongly correlated states
has created a platform in which the relation between the properties of the electronic bands and 
the nature of the correlated states can be studied in unprecedented ways. The reason is that
these systems allow extraordinary control of the electronic bands' properties, for example
by varying the relative
twist angle between the layers forming the system. In particular, in twisted bilayers the low energy bands can
be tuned to be very flat and with a nontrivial quantum metric. This allows the quantitative
and experimental exploration of the relation between the metric of Bloch quantum states
and the properties of correlated states. In this work we first review the general connection
between quantum metric and the properties of correlated states that break a continuous symmetry.
We then discuss the specific case when the correlated state is a superfluid and show 
how the quantum metric is related to the superfluid stiffness.
To exemplify such relation we show results for the case of superconductivity
in magic angle twisted bilayer graphene.
We conclude by discussing possible research directions to further elucidate
the connection between quantum metric and correlated states' properties. 
\end{abstract}


\maketitle


\section{Introduction}
One of the most exciting developments of the past few years in condensed matter physics has been
the ability of experimentalists to realize two-dimensional (2D) ``twisted bilayers''~\cite{li2010observation} 
and observe the establishment in these
systems of strongly correlated electronic states~\cite{Kim2017b,Cao2018,Cao2018c,Yankowitz2019a,Chen2019,Lu2019b,Choi2019,Sharpe2019,Polshyn2019,Codecido2019,Liu2020a,Shen2020,Chen2020,Cao2020,Wang2020}.
These systems are formed by two 2D crystals stacked with a relative twist angle $\theta$.
Twisted bilayer graphene (TBLG), formed by two graphene layers, so far, has been the most studied twisted bilayer system.
The feat that experimentalists have been able to accomplish is to control $\theta$ with high precision and
tune it to particular, ``magic'', values ($\theta_M$) for which the bands of the system are almost completely flat~\cite{dossantos2007,morell2010,Bistritzer2011}.
It is for this magic values of $\theta$ that the system exhibits a very rich phase diagram with
strongly correlated phases, including a superconducting phase for which the ratio between the critical temperature, $T_c$,
and the Fermi temperature, $T_F$, ranges between 0.04 and 0.1, depending on the doping~\cite{Cao2018}.
%
%
The value of $T_c/T_F\approx 0.1$ is much larger than the one for conventional BCS superconductors, and 
implies that to understand the origin of superconductivity in MATBLG weak coupling theory is not sufficient.
Such value is also larger than in most unconventional superconductors~\cite{Cao2018}, in particular
high $T_c$ cuprates.
%

One very interesting aspect of magic angle twisted bilayer graphene (MATBLG) is the non-trivial geometry of its quantum states.
As a consequence MATBLG is a new, highly tunable, platform in which the connection between strong correlations and quantum states' geometry
can be explored in detail both theoretically and experimentally. This allows to significantly advance our understanding
of the relation between the metric of quantum states, the conditions necessary for the establishment and stability of strongly
correlated states, and the properties of these states.

For the past fifteen years the geometry of quantum states 
has been at the center of some of the most interesting discoveries in condensed matter physics.
The geometry of a manifold of quantum states is encoded by the ``quantum geometric tensor'', \GG~\cite{Provost1980a,Page1987,Bengtsson2017,Cheng2010}.
\GG has both a real and an imaginary part.
The imaginary part of \GG corresponds to the Berry curvature~\cite{Berry1984}.
In the past few years many interesting developments
in condensed matter physics have arisen by a careful treatment of the Berry curvature.
Exemplary are the the discovery of topological insulators (TIs) and superconductors~\cite{Read2000,schnyder2008,hasan2010,qi2011rmp,chiu2016},
Weyl and Dirac semimetals (SMs)~\cite{Wehling2014,Vafek2014,Armitage2018,Burkov2018}, and, more recently,
higher order topological materials~\cite{Benalcazar2017a,Schindler2018,Song2017,Khalaf2018,Ezawa2018,Trifunovic2019,Ghorashi2019,Ghorashi2019d,FangCano2020}.
At the same time, it is interesting to notice  how much less attention the real
part of \GG, \ReGG, the ``quantum metric'', has received compared to its imaginary part.
This is in great part due to the difficulty to measure physical quantities related to \ReGG.
However, the connection between quantum metric and the properties of collective ground states breaking a continuous symmetry,
and the availability of a system like MATBLG, have opened a new avenue to understand how \ReGG
can affect the macroscopic properties of quantum systems.

In this work we briefly review the recent progress in the understanding of the relation
between quantum metric and the properties of correlated states of 2D systems.
In section~\ref{sec:general} we present the formalism describing in general terms the relation
between quantum metric and correlated states,
in section~\ref{sec:superfluid} we discuss the case when the correlated state is a superconductor,
in section~\ref{sec:tblg} we review some of the recent results for MATBLG,
and finally in section~\ref{sec:outlook} we summarize the current status of our understanding of
the topic and possible developments in the near future.
%


\section{Quantum metric and properties of many-body systems}
\label{sec:general}
Quantum mechanical states are represented by rays in a complex Hilbert space.
For a given quantum system, therefore, the space of physical states is not the Hilbert space
\hh, but the projective Hilbert space \ph. 
The projective Hilbert space \ph is the space formed by rays in the Hilbert space \hh, where each ray is the set of vectors in \hh of unit norm that differ only by multiplication by phase factors.  For a Hilbert space of dimension $n$, 
\ph is the complex projective space $\mathbb{C}P^{n-1}$ formed by the lines through the origin of a complex Euclidean space. The inner-product of \hh
endows \ph with the structure of a K\"ahler manifold, i.e. a manifold with a 
a proper metric tensor~\cite{Kibble1979,Ashtekar1999}.
\ph can be parametrized by an element $\lambda$
of a space $\cal{V}$ (that itself can be a manifold) like real space, or momentum space.
In the remainder we assume the space $\cal{V}$ to be the momentum space
with elements identified by the momentum wave-vector $\kk$.

The inner product of \hh leads to the natural definition of
the distance $d s^2$
between two vectors $\ket{\psi(\kk)}$, $\ket{\psi(\kk+d\kk)}$
with infinitesimally close
momenta, $\kk$, $\kk+d\kk$:
$ds^2=\braket{\partial_\mu\psi}{\partial_\nu\psi}dk^\mu dk^\nu$,
where $\partial_\mu \df \partial/\partial k_\mu$.
Given that quantum states are represented by elements of \ph, not \hh,
the expression of $ds^2$ is not the proper distance between two quantum states with
infinitesimally close momenta. This is also reflected by the fact that
$\braket{\partial_\mu\psi}{\partial_\nu\psi}$, in general, is not gauge invariant.
The proper distance between quantum states can be obtained by redefining $ds^2$ to
remove the effects of a gauge transformation~\cite{Provost1980a,Page1987,Bengtsson2017,Cheng2010}.
This leads to the expression:
\begin{align}
 ds^2 &= Q_{\mu\nu} dk^\mu dk^\nu; \hspace{0.15cm} \\
 Q_{\mu\nu}&\df \braket{\partial_\mu\psi}{\partial_\nu\psi}-\braket{\partial_\mu\psi}{\psi}\braket{\psi}{\partial_\nu\psi} \label{eq:Q} \\
 B_{\mu\nu}&\df {\rm Im}[Q_{\mu\nu}] \label{eq:Q-Bmunu} \\
 g_{\mu\nu}&\df {\rm Re}[Q_{\mu\nu}] \label{eq:Q-gmunu} 
\end{align}
where we have introduced the {\em quantum geometric tensor} $Q_{\mu\nu}$.
$Q_{\mu\nu}$ is gauge invariant. Its imaginary part is the Berry curvature, $B_{\mu\nu}$, and
is completely antisymmetric and therefore does not contribute to $ds^2$.
Its real part, $g_{\mu\nu}$,
is the Fubini-Study quantum metric~\cite{Fubini1904,Study1905}.
It is interesting to point out that
the Fubini-Study metric is the unique Riemannian metric on \ph that is invariant under the action 
of unitary transformations ($U(n)$) on $CP^{n-1}$.
%
The quantum geometric tensor $Q_{\mu\nu}$ is positive semidefinite~\cite{Provost1980a}. This fact implies
the following two inequalities~\cite{Roy2014}:
\begin{align}
 \det g_{\mu\nu}  &\geq |B_{\mu\nu}|^2, \label{eq:bound01} \\
 {\rm Tr} g_{\mu\nu}&\geq 2|B_{\mu\nu}| \label{eq:bound02}
\end{align}
%
%

It is possible to generalize the definition of $Q_{\mu\nu}$
to the non-Abelian case~\cite{Ma2010},
in analogy to the non-Abelian generalization of the Berry curvature~\cite{Wilczek1984}.
In this generalization one takes into account that at the degeneracy points quantum states related by a rotation
in the subspace spanned by the degenerate eigenstates are equivalent. By properly projecting 
$\braket{\partial_\mu\psi}{\partial_\nu\psi}$
one obtains the  gauge invariant ``non-Abelian'' quantum metric.

The impact of the study of the effects of the Berry curvature ${\rm Im}[Q_{\mu\nu}]$ on the properties
of quantum systems cannot be overstated. 
Just in the context of condensed matter systems the Berry curvature, and associated Berry phase~\cite{Berry1984}, 
greatly impacted the understanding of the quantum Hall effect, the anomalous Hall effect, orbital magnetism~\cite{Xiao2010b} and it lead to 
the discovery of topological materials~\cite{hasan2010,bernevig2013topological}, and Weyl semimetals~\cite{Armitage2018}.
By contrast the effect of ${\rm Re}[Q_{\mu\nu}]$
has so far been much less studied.
\gm has been shown to be connected to the
Hall viscosity~\cite{Avron1995,Read2009,Haldane2009,Read2011,Hoyos2012,Bradlyn2012,Haldane2015,Shapourian2015,Holder2019a},
a quantity that is difficult to measure~\cite{Hughes2011,Delacretaz2017a,Scaffidi2017a,Pellegrino2017,Berdyugin2019,Rao2020,Gianfrate2020}.
For a perfect conductor the longitudinal electric conductivity $\sigma_{xx}(\omega)$ as a function of frequency $\omega$
has a delta function $D\delta({\omega})$, where $D$ is the Drude weight.
The Drude weight has also been shown to be connected to the quantum metric \gm,~\cite{Resta2011b,Resta2018a,Marrazzo2019,Bellomia2020}.
Such connection, however, is also difficult to ascertain experimentally given
that at finite temperature, or in the presence of any amount of disorder,
$\sigma_{xx}(\omega)$ does not have a Dirac's delta for $\omega=0$ and therefore $D=0$.

The experimental challenges to verify the relation between \gm, the Hall viscosity, and $D$ are likely an important reason
for the fact that much less research activity has been focused on the study of the effects of
the quantum metric than on the study of the effects of the Berry curvature.
%
%
Recently, however, novel 
connections~\cite{Roy2014,Jackson2015,Gao2015z,Raoux2015,Piechon2016,Rhim2020,Srivastava2015,Smith2021}
have been made between the quantum metric and 
properties of electronic systems. In Refs.~\cite{Roy2014,Jackson2015} the quantum metric
of a fractional Chern insulator~\cite{Regnault2011}
has been bee shown to be related to the stability of the fractional quantum Hall (FQH) phase of these systems.
In particular it was shown that for a fractional Chern insulator band $j$
the trace ${\rm Tr}[g^{(j)}_{\mu\nu}-|B_{\mu\nu}|]$ is correlated to the gap of the
FQH-like phase~\cite{Jackson2015}.
It is also known that the magnetic susceptibility of a periodic multi orbital electron system
depends on the metric properties of the quantum states~\cite{Blount1962,Gao2015z,Raoux2015}.
In Ref.~\cite{Piechon2016} this connection has been made more explicit for the case of two-band models.
The metric tensor of a singular 2D flat band~\cite{Rhim2019}, 
i.e. a flat band with a crossing point with a dispersive band, 
has also been shown to be connected to the {\em energy spread} of the Landau levels
arising from the singular 2D flat band in the presence of a magnetic field~\cite{Rhim2020}.

For systems in which the interactions induce a collective ground state that breaks a $U(1)$ symmetry
it has become apparent that the quantum metric is connected to the phase {\em stiffness}, \rhos, of
the collective ground state.
This can be seen considering that in this case the effective Ginzburg-Landau action
describing the low energy physics of the collective ground state has a term of the form
\begin{equation}
 S = \beta\frac{1}{2}\int d\rr \rho^{(s)}  |\nabla\psi|^2
 \label{eq:S}
\end{equation}
where $\psi=\psi_0 e^{i\phi}$ is the complex order parameter describing the ground state,
$\psi_0$ being the amplitude and $\phi$ the phase parametrizing $U(1)$,
and $\beta=1/(k_B T)$, $T$ being the temperature and $k_B$ the Boltzmann constant.
To simplify the notation in Eq.~\ceq{eq:S} we have assumed
the stiffness to be diagonal and isotropic $\rhosm=\rho^{(s)}\delta_{\mu\nu}$.
We can then introduce a gauge field ${\bf A}_{\rm eff}$ associated to the $U(1)$ charge $\tilde e$.
In the presence of ${\bf A}_{\rm eff}$ the gradient in Eq.~\ceq{eq:S} must be replaced by
the gauge covariant gradient $\nabla - i\tilde e {\bf A}_{\rm eff}$ from which we get mix terms 
of the form $-i\tilde e {\bf A}_{\rm eff}\nabla$ that describe the coupling of the system to the 
field ${\bf A}_{\rm eff}$. 
From this we can see that 
that the current operator ${\bf j}$
coupling to ${\bf A}_{\rm eff}$ is $\sim \tilde e\nabla_{\xx}$,
and that $\rho_s$ must be related to the strength of the current-current response ($K$) of the system
to the probing field ${\bf A}_{\rm eff}$. 
This is completely analogous
to the case of a superconductor, discussed in the next section, in which the connection
between the metric of the quantum states and \rhos is shown explicitly.
%
%
This connection was first shown explicitly for simple cases in superconductors~\cite{Peotta2015,Julku2016,Liang2017}
and for flat ferromagnetic states in systems with flat bands~\cite{Liang2017a}.

Among all the types of condensed matter systems in which the ground states spontaneously break a $U(1)$ symmetry
two are particularly important and common:
ferromagnets (FMs) and superconductors (SCs).
For both classes of systems \ReGG can play an essential role in determining the properties
of the collective ground state. For magnetic systems \ReGG enters the expression of the spin-stiffness, \rhosspin,
for superconductors it contributes to the superfluid stiffness, \rhos, or, equivalently, the superfluid weight \ds.
\rhosspin and \rhos can be measured and are not affected by small amounts of disorder
and so their relationship to \ReGG can be verified experimentally.

For 2D systems for which the ground state spontaneously breaks a U(1) symmetry, \rhos
governs the Berezinskii-Kosterlitz-Thouless~\cite{Berezinski1971,Kosterlitz1973}
(BKT) transition, in particular it fixes the value of the temperature, \tkt,
at which the transition takes place. For an isotropic system
$\rhosm = \rho^{\rm (s)}\delta_{\mu\nu}$ and $\rho^{\rm (s)}$ fixes \tkt via the relation~\cite{Kosterlitz1973}:
\beq
 k_B T_{KT} =\frac{\pi}{2}\rho^{(s)}(T_{KT}).
 \label{eq:tkt}
\enq 
As we discuss in the following two sections, equation~\ceq{eq:tkt} can be used to estimate the value of $\rho_s$ in 2D systems.

For a multi-orbital system \rhos has a contribution due to the curvature of the bands, the so called ``conventional'' contribution, \rhosconv,
and a contribution due to \ReGG, the so called ``geometric'' contribution, \rhosg.
\ReGG can be different from zero only for multiband systems. 
It is therefore clear that the geometric contribution to \rhos can be dominant in multi-orbital systems with flat bands.
This is precisely the situation in MATBLG: the effective moir\'e lattice of MATBG has a multiband spectrum
with the lowest energy bands, the ones that participate in the formation of collective ground states
such as superconducting and ferromagnetic states~\cite{Andrei2020,Andrei2021,Lin2021}, extremely flat.
The advent of systems like MATBLG has then greatly increased our ability to
study and understand the relation between the metric of quantum states and 
the macroscopic properties of collective ground states.

\section{Quantum metric and superfluid stiffness}
\label{sec:superfluid}

To exemplify in concrete terms the connection between the quantum metric
and the stiffness of a ground state breaking a $U(1)$ symmetry we consider
the case of a superconductor. 
For the linear 
current response to an external vector potential,
in momentum and frequency space we have
\beq
 j_\mu(\kk,\omega) = K_{\mu\nu}(\kk,\omega) A_\nu(\kk,\omega)
 \label{eq:jlin}
\enq
where $j_\mu(\kk,\omega)$, $A_\mu(\kk,\omega)$, and  $K_{\mu\nu}(\kk,\omega)$
are the Fourier amplitude  with wave vector $\kk$ and frequency $\omega$ of the $\mu$ component of the current density, 
the $\mu$ component of the vector potential ${\bf A}$,
and of the $\mu\nu$ component of the current-current response function, respectively.
The superfluid weight, \ds is the tensor that relates, within the linear approximation,
$j_\mu$ to the $\nu$ component of a static ($\omega=0$) transverse vector potential, $\kk\cdot\aaa=0$,
in the limit $\kk\to 0$. Denoting by $k_\parallel$, $k_\perp$, the components of $\kk$ parallel and perpendicular to $\aaa$, respectively, 
we have~\cite{Scalapino1992,Scalapino1993}:
\beq
  D^{\rm {(s)}}_{\mu\nu} \equiv -\lim_{k_\perp\to 0}K_{\mu\nu}(k_\parallel = 0,\omega=0).
 \label{eq:ds}
\enq
By combining Eqs.~\ceq{eq:jlin},~\ceq{eq:ds} we obtain London's equation
\beq
  \lim_{k_\perp\to 0} j_\mu(k_\parallel = 0,\omega=0) = -D^{\rm {(s)}}_{\mu\nu} \lim_{k_\perp\to 0}A_\nu(k_\parallel = 0,\omega=0)
  \label{eq:london}
\enq 
that captures the key features, such as the Meissner effect, of the superconducting state.
\rhos is directly proportional to $D^{\rm {(s)}}_{\mu\nu}$:
\beq
  \rhosm = \frac{\hbar^2}{e^2}D^{\rm {(s)}}_{\mu\nu}
\enq

Notice that Eq.~\ceq{eq:london} was obtained requiring $\omega=0$, $k_\parallel = 0$, and then taking the limit $k_\perp\to 0$.
As a consequence Eq.~\ceq{eq:london} cannot be used to relate a time-dependent current to a time-dependent vector potential.
This can only be done by allowing $\omega\neq 0$ when calculating $K_{\mu\nu}(\kk,\omega)$.
The value of $K_{\mu\nu}(\kk,\omega)$ in the limit  $(\kk=0, \omega\to 0)$ is proportional to the Drude weight~\cite{Scalapino1992,Scalapino1993}
(see Sec.~\ref{sec:general}).

For an isolated parabolic band, at zero temperature, $\rhosm =\hbar^2 (n/m^*)\delta_{\mu\nu}$~\cite{Scalapino1992,Scalapino1993},
where $n$ is the electron density, and $m^*$ is the effective mass of the band.
This conventional result 
would lead us to the conclusion that for systems like MATBLG, for which $m^*\to\infty$, \rhos should be very small
so that the hallmark signatures of superconductivity such as the Meissner effect (for 3D systems)
should be extremely weak. This is in contrast with the experimental observations and shows that
the conventional expression for \rhos obtained for a single parabolic band is not general enough.

%
%
For the case of a multi-band system we need to derive the expression of \rhos from the general
expression of $K_{\mu\nu}(\kk,\omega)$. Using the Kubo formula we have:
\begin{align}
 K_{\mu\nu}(\kk,\omega) = \langle T_{\mu\nu}\rangle + \langle \chi_{\mu\nu}^p(\kk,\omega)\rangle
	 \label{eq:K}
\end{align}
where $T_{\mu\nu}$ is the diamagnetic current operator
\begin{equation}
 T_{\mu\nu} = \sum_{\sigma}\int \frac{d\kk}{(2\pi)^d} c^\dagger_{\kk\sigma}\partial_\mu\partial_\nu H(\kk,\sigma)c_{\kk\sigma},
 \label{eq:T}
\end{equation}
and
\beq
 \chi_{\mu\nu}^p(\kk,\omega) = -i \int_0^\infty dt e^{i\omega^+ t}\langle[j^p_\mu(\kk,t),j_\nu^p(-\kk,0)]\rangle
 \label{eq:chiP}
\enq
is the time Fourier transform of the correlator of the paramagnetic current operator
\begin{equation}
 j_{\mu}^p(\kk) = \sum_{\sigma}\int \frac{d\kk}{(2\pi)^d} c^\dagger_{\kk'\sigma}\partial_\mu H(\kk'+\kk/2,\sigma)c_{\kk'+\kk\sigma}.
 \label{eq:jp}
\end{equation}
The angle brackets denote expectation values over the ground state, and $[,]$ is the commutator.
In Eq.~\ceq{eq:T},~\ceq{eq:jp} $c^\dagger_{\kk'\sigma}$ ($c_{\kk'\sigma}$) is the creation (annihilation) operator
for an electron with momentum $\kk$ and spin $\sigma$, and $d$ is the dimensionality of the system. 
$H$ is the matrix Hamiltonian describing the system
expressed in the basis used for the creation annihilation operators (spin-momentum basis).

A superconductor can be described in general by a Bogolyubov de Gennes Hamiltonian ${\cal H}_{\rm BdG}$
of the form:
\beq
   {\cal H}_{\rm BdG} = 
  (\psi^\dagger_T \psi_B) 
  \hbdg 
  \begin{pmatrix} \psi_T \\ \psi^\dagger_B \end{pmatrix}, \hspace{0.25cm}
  \hbdg =  \begin{pmatrix}
   H_T & \hat\Delta\\
   \hat\Delta^\dagger & -H_B
  \end{pmatrix}
  \label{eq:Hbdg}
\enq
where $\psi^\dagger_T$, $\psi^\dagger_B$ ($\psi_T$, $\psi_B$) are the creation (annihilation) spinor operators
for the states, described in the normal phase by the matrix Hamiltonians $H_T$, $H_B$, respectively, 
that pair to form the condensate characterized by the pairing matrix $\hat\Delta$.
Using the expression of $\hbdg$ given in Eq.~\ceq{eq:Hbdg}, for $\langle T_{\mu\nu}\rangle$, in the Matsubara formalism, we obtain:
\beq
 \langle T_{\mu\nu}\rangle = \frac{1}{\beta}\int \frac{d\kk}{(2\pi)^d} \sum_{\omega_n} \tr[\partial_\mu\partial_\nu \hbdg G(i\omega_n,\kk)]
 \label{eq:T02}
\enq
where $\omega_n=\pi k_B T(2n +1)$, with $n\in \mathbb Z$, are the fermionic 
Matsubara frequencies and
\beq
 G(i\omega_n,\kk)=[i\omega_n - \hbdg]^{-1}= \sum_{j}\frac{\ket{\psi_j(\kk)}\bra{\psi_j(\kk)}}{i\omega_n - E_j(\kk)}
 \label{eq:G}
\enq
is the retarded Green's function. 
In Eq.~\ceq{eq:G} $E_j$ and $\ket{\psi_j(\kk)}$
are the eignenvalues and eigenvectors, respectively, of $\hbdg$.
By performing the integration over $\kk$ by parts, and considering that, from the definition of $G$,
$\partial\mu G = -G^2\partial_\mu \hbdg$, we can rewrite Eq.~\ceq{eq:T02} in the form:
\beq
 \langle T_{\mu\nu}\rangle = \frac{1}{\beta}\int \frac{d\kk}{(2\pi)^d} \sum_{\omega_n} \tr[\partial_\mu\hbdg G^2(i\omega_n,\kk)\partial_\nu\hbdg].
 \label{eq:T03}
\enq
Similarly for the contribution arising from the paramagnetic currents we obtain:
\begin{align}
 \langle \chi_{\mu\nu}^p(\kk,i&\Omega_m)\rangle =
 \frac{1}{\beta}\int \frac{d\kk'}{(2\pi)^d} \sum_{\omega_n} \tr[G(i\omega_n,\kk') \nonumber \\
 &\partial\nu\hbdg(\kk'+\kk/2)\tau_z G(i\omega_n+i\Omega_m,\kk'+\kk) \nonumber \\
 &                                         \partial\mu\hbdg(\kk'+\kk/2)\tau_z].
 \label{eq:Kpm}                                           
\end{align}
where $\Omega_m=2\pi m k_B T$ ($m\in\mathbb Z)$ are the bosonic Matsubara frequencies, and $\tau_z$ is the $z$-Pauli matrix.
Combining Eqs.~\ceq{eq:K},~\ceq{eq:G}, \ceq{eq:T03}, and \ceq{eq:Kpm}, after summing over the fermionic Matsubara frequencies,
in the limit $i\Omega_m=0$, $\kk\to 0$, we obtain~\cite{Liang2017}:
\begin{align}
 \rhosm = &\sum_{i,j}\int \frac{d\kk}{(2\pi)^d} \frac{n_F(E_i)-n_F(E_j)}{E_j-E_i} \nonumber \\
 &\left[\bra{\psi_i}\partial_\mu H_{\rm BdG}\ket{\psi_j} 
 \bra{\psi_j}\partial_\nu H_{\rm BdG}\ket{\psi_i}\right. \nonumber \\ 
 &\left. -\bra{\psi_i}\partial_\mu H_{\rm BdG}\tau_z\ket{\psi_j} 
 \bra{\psi_j}\tau_z\partial_\nu H_{\rm BdG}\ket{\psi_i} 
 \right]
 \label{eq:ds_gen}
\end{align}
where $n_F(E)$ is the Fermi-Dirac function.

Equation~\ceq{eq:ds_gen} can be used to show the connection between \rhos and the quantum metric
of the Bloch states.
The origin of such connection can be understood by considering that in general, for a generic Hamiltonian $H$, 
the expectation values $\bra{\psi_i}\partial_\mu H\ket{\psi_j}$ 
of the velocity operator $\partial_\mu H$
have an anomalous contribution proportional to $\braket{\psi_i}{\partial_\mu\psi_j}$,
and that therefore the terms 
$\bra{\psi_i}\partial_\mu H_{\rm BdG}\ket{\psi_j}\bra{\psi_j}\partial_\nu H_{\rm BdG}\ket{\psi_i}$
in Eq.~\ceq{eq:ds_gen} 
give rise to terms of the form $\braket{\partial_\mu\psi_i}{\partial_\nu\psi_i}$
that, as shown above, Eq.~\ceq{eq:Q},~\ceq{eq:Q-gmunu}, enter the expression of the quantum metric.
We call the part of \rhos arising from these terms the ``geometric part'', \rhosg, of \rhos.

We can explicitly separate the contribution to \rhos arising from the metric of the quantum states
from the conventional one, arising from terms proportional to the derivatives of the eigenvalues with respect to $\kk$.
Let $\{\emt\}$ ($\{\emb\}$), $\{\mt\}$ ($\{\mb\}$) be the eigenvalues and eigenstates, respectively, of $H_T$ ($H_B$).
The Hilbert space for $\hbdg$ is given by the direct sum of the Hilbert spaces $\hs_T$ of $H_T$
and $\hs_B$ of $H_B$.
Any eigenstate $\ket{\psi_i}$ of $\hbdg$ can be written as $(\ket{\psi^T_{i}},\ket{\psi^B_{i}})$
with $(\ket{\psi^T_{i}}\in \hs_T$, and $(\ket{\psi^B_{i}}\in \hs_B$.
Assuming $\hat\Delta$ to be independent of $\kk$, following~\cite{Liang2017}, we can rewrite Eq.~\ceq{eq:ds_gen}
to identify the contribution to \rhos arising from the quantum metric of the $\mt$, $\mb$ 
states, i.e. the quantum metric of the bands in the normal phase. To do this 
we start by rewriting the expectation values $\bra{\psi_i}\partial_\mu \hbdg \ket{\psi_j}$
in terms of the $\mt$, $\mb$ states
\begin{align}
 \bra{\psi_i}\partial_\mu \hbdg \ket{\psi_j} = \sum_{\substack{m_T,m_B\\n_T,n_B}}
                                               [&c^T_{i,m_T} J^T_{\mu,m_T,n_T} c^T_{n_T,j} - \nonumber \\
                                                &c^B_{i,m_B} J^B_{\mu,m_B,n_B} c^B_{n_B,j}]
  \label{eq:psim}                                          
\end{align}
where
\begin{align}
  c^X_{i,m_X} &= \braket{\psi^X_i}{m_X}; \label{eq:cxm} \\
  J^X_{\mu,m_X,n_X}& = \bra{m_X}\partial_\mu \hbdg \ket{n_X}.
  \label{eq:jxm}
\end{align}
and $X=(T, B)$. To simplify the notation in Eqs.~\ceq{eq:psim}-\ceq{eq:jxm} we do not
show explicitly the dependence of the quantities on the momentum $\kk$.
Using Eqs.~\ceq{eq:psim}-\ceq{eq:jxm} we can rewrite Eq.~\ceq{eq:ds_gen} in the form
\begin{align}
 \rho^{(s)}_{\mu\mu} = &-4\!\!\!\!\sum_{\substack{m_T, n_T\\ p_B, q_B,i,j}}\int\frac{d\kk'}{(2\pi)^d}
 		 {\rm Re} \left[\frac{n_F(E_i)-n_F(E_j)}{E_j-E_i} \right. \nonumber \\
        & c^T_{i,m_T} (c^T_{j,n_T})^*c^B_{j,p_B} (c^B_{i,q_B})^* 
          J^T_{\mu,m_T,n_T}J^B_{\mu,p_B,q_B} ]  
         \label{eq:ds_con_geo}   
\end{align}
The current expectation values $J^X_{\mu,m_X,n_X}$ can be written as
\beq
 J^X_{\mu,m_X,n_X} = \partial_\mu \emx \delta_{m_X,n_X} + (\enx - \emx)\braket{m_X}{\partial_\mu n_X}.
 \label{eq:J}
\enq
Equation~\ceq{eq:J} shows that $J^X_{\mu,m_X,n_X}$ has a "conventional" contribution proportional to $\partial_\mu \emx$,
and a contribution, the second term in Eq.~\ceq{eq:J}, related to the geometry of the quantum states.
Combining Eq.~\ceq{eq:ds_con_geo} and Eq.~\ceq{eq:J} we can then identify three contributions to \rhos=\rhosa + \rhosb + \rhosc,
\begin{align}
 \rhosam = & -\!\!4\!\!\!\!\!\!\!\sum_{\substack{m_T, n_T\\ p_B, q_B,i,j}}\!\!\!\int\!\!\!\frac{d\kk}{(2\pi)^d}{\rm Re}\!
             \left[ C_{m_T n_T}^{p_B q_B} \partial_\mu\emt\partial_\nu\eqb\delta_{m_Tn_T}\delta_{p_Bq_B} \right] 
             \label{eq:rho-1}  \\
 \rhosbm = & -\!\!4\!\!\!\!\!\!\!\sum_{\substack{m_T, n_T\\ p_B, q_B,i,j}}\!\!\!\int\!\!\!\frac{d\kk}{(2\pi)^d}{\rm Re}
             \left[C_{m_T n_T}^{p_B q_B}[\partial_\mu\emt\delta_{m_T n_T}\right. \nonumber \\
              &(\eqb - \epb)\braket{p_B}{\partial_\mu q_B}  \nonumber \\
              &+\left. \partial_\nu\epb\delta_{p_Bq_B}(\ent - \emt)\braket{m_T}{\partial_\nu n_T}]\right]
              \label{eq:rho-2}  \\
 \rhoscm = & -\!\!4\!\!\!\!\!\!\!\!\sum_{\substack{m_T\neq n_T\\ p_B\neq q_B,i,j}}\!\!\!\int\!\!\!\frac{d\kk}{(2\pi)^d}{\rm Re}
             \left[C_{m_T n_T}^{p_B q_B}
             [ (\eqb - \epb)(\ent - \emt) \right. \nonumber \\ 
             &\hspace{2cm}\left. \braket{p_B}{\partial_\mu q_B}\braket{m_T}{\partial_\nu n_T}]\right]. 
 \label{eq:rho-3}                                      
\end{align}
where 
\beq
 C_{m_T n_T}^{p_Bq_B} \equiv \sum_{ij}\frac{n_F(E_i)-n_F(E_j)}{E_j-E_i} c^T_{i,m_T} (c^T_{j,n_T})^*c^B_{j,p_B} (c^B_{i,q_B})^*.
\enq 
\rhosa is the conventional contribution to \rhos.
\rhosb is a "mixed" contribution: it depends in part on the properties of the band dispersion, as the conventional
part, and in part on the geometry of the quantum states. For systems with particle-hole symmetry this term is negligible.
\rhosc has only terms proportional to
$\braket{m_T}{\partial_\nu n_T}$, i.e. terms that depend on the metric properties of the quantum states; there are no terms proportional to the gradient of the eigenvalues with respect to $\kk$. 
For this reason, it is natural to identify \rhosc as the dominant geometric term.
\rhos and \rhosa are gauge invariant and therefore the combination \rhosb+\rhosc is also gauge invariant.
For this reason it is useful at times to separate \rhos in the two terms: \rhosa that only depends on the 
bands' dispersion, and \rhosb+\rhosc that is mostly given by the metric properties of the Bloch states.
In the remainder, considering that we mostly focus on superconducting systems
for which \rhosb is negligible, we identify \rhosc as the geometric part, \rhosg, of \rhos.

The expression of \rhosg$\equiv$\rhosc given by Eq.~\ceq{eq:rho-3}, as long as the order parameter is independent of momentum, 
is quite general and therefore shows the general nature of the connection
between quantum metric and superfluid density. It is fairly straightforward to write a similar
equations for the spin stiffness of a XY ferromagnet or the pseudo-spin stiffness
of an XY orbital-ferromagnet, i.e., a state in which the degree of freedom ordering is not the
spin but an orbital degree of freedom, situation that appears to be very relevant for
systems like MATBLG~\cite{Sharpe2019,Xie2020,Serlin2020a,Wu2021a}.

It is instructive to see how Eqs.~\ceq{eq:rho-1},~\ceq{eq:rho-3} simplify 
when the chemical potential lies within a well
isolated band, $j$. In this case,
neglecting terms of order $1/\Gamma_{ij}$, where $\{\Gamma_{ij} \}$ are the gaps 
between band $j$ and the other bands, and assuming the pairing matrix to be proportional to the identity with amplitude $\Delta$,
we can obtain a direct relation between 
\rhosg and  the quantum metric $g_{\mu\nu}^{(j)}$ of band $j$ when time-reversal symmetry is preserved and
the superconducting order parameter, in addition to being $\kk$-independent,
only has intraband terms. In this case we have~\cite{Liang2017}:
\begin{align}
  \rhosm =&\int\!\!\!\frac{d\kk}{(2\pi)^d}\left[ 2\frac{\partial n_F(E_j)}{\partial E_j}+\frac{1-2n_F(E_j)}{E_j} \right]
         \frac{\Delta^2}{E_j^2}\partial_\mu\eps_j\partial_\nu\eps_j + \nonumber \\ 
         &2\Delta^2\int\!\!\!\frac{d\kk}{(2\pi)^d}        
         \frac{1-2n_F(E_j)}{E_j}
         g_{\mu\nu}^{(j)} 
 \label{eq:ds_gmunu}
\end{align}
where $\kk$ is the momentum.
The last term in Eq~\ceq{eq:ds_gmunu} is the geometric part of \rhos that, in this simple case,
is related in a very direct way to the quantum metric $g_{\mu\nu}^{(j)}$ of the isolated band. 

Using the expression above, and 
the inequality~\ceq{eq:bound02}
{\em for the case of an isolated band} we can provide a bound for the geometric part of \rhos~\cite{Peotta2015,Liang2017}:
\begin{equation}
 \rho^{\rm (s, geo)_{\mu\nu}}\geq 2\Delta^2\int\!\!\!\frac{d\kk}{(2\pi)^d} \frac{1-2n_F(E_j)}{E_j} |B_{\mu\nu}|^2.
 \label{eq:bound1}
\end{equation}
This result shows that for bands with large Berry curvature the geometric contribution to \rhos is large.
It is important to point out that Eq.~\ceq{eq:bound1} only provides a lower bound given that it is possible
to have situations in which $g_{\mu\nu}\neq 0$ even if the Berry curvature is zero~\cite{Piechon2016}.

In 2D, for the case in which the isolated band, is flat, i.e. having a bandwidth much smaller than 
the $\Gamma_{ij}$ gaps, and {\em non degenerate}, \rhos is only given by the geometric part and can be written in the form~\cite{Peotta2015}:
\begin{equation}
 \rhosm = 2\Delta\sqrt{\nu(1-\nu)}\int\frac{d\kk}{(2\pi)^2}g_{\mu\nu}(\kk).
 \label{eq:2Dflat}
\end{equation}
where $\nu$ is the filling fraction of the flat band.
In this case we have that $(1/2\pi)\int d\kk B_{\mu\nu}=\varepsilon_{\mu\nu}C$, where 
$\varepsilon_{\mu\nu}$ is the $2\times 2$ Levi-Civita tensor and $C$ is the Chern number of the isolated band.
Using inequality~\ceq{eq:bound01} we obtain $\det(\int d\kk g_{\mu\nu})\geq\det(d\kk \int |B_{\mu\nu}|^2=C^2$ and then,
for an isotropic system~\cite{Peotta2015}:
\begin{equation}
 \rho^{\rm (s)}\geq \frac{\Delta}{\pi}\sqrt{\nu(1-\nu)}|C|.
 \label{eq:bound2}
\end{equation}

In general, when the 2D flat band has degenerate points it might not be possible to find a lower bound for \rhos=\rhosg, however, 
this can be done for the case relevant to MATBLG in which the two low-energy 2D flat band have degeneracy points and 
$C_{2z}{\cal T}$ symmetry, $C_{2z}$ being the twofold
rotation around the $z$-axis perpendicular to the 2D plane to which the quantum states are confined, and ${\cal T}$ the time-reversal
operator~\cite{Xie2019}. Given the degeneracy of the bands it is necessary to consider the non-Abelian generalization
of the expression of \GG. It can be shown that the $C_{2z}T$ symmetry constrains the non-Abelian Berry curvature
to the form~\cite{Song2019,Ahn2019} 
$B_{xy}=-b_{xy}(\kk)\sigma_2$
with $(1/2\pi)\int d\kk b_{xy}=e_2$, where $e_2$ is the Wilson loop winding number~\cite{Song2019}, or "Euler's class"~\cite{Ahn2019}, 
of the two bands.
In this case, assuming the pairing $\Delta$ is non vanishing only for the low-energy, two-fold degenerate, band, 
and using again inequality~\ceq{eq:bound02},
we have that $\rho^{\rm (s)}$ has the lower bound~\cite{Xie2019}
\begin{equation}
 \rho^{\rm (s)} \geq \frac{\Delta}{\pi}\sqrt{\nu(1-\nu)}|e_2|.
 \label{eq:bound3}
\end{equation}
For the specific case of TBLG $e_2=1$ so that, taking into account the spin and valley degeneracy,
we obtain~\cite{Xie2019}
\begin{equation}
 \rho^{\rm (s)} \geq  4\frac{\Delta}{\pi}\sqrt{\nu(1-\nu)}.
 \label{eq:bound3tblg}
\end{equation}
Inequalities \ceq{eq:bound2}, \ceq{eq:bound3} show how the topological invariants of the bands can be used to
obtain lower bounds for \rhos in flat-band systems.

For a superconductor \rhos determines the phase-stiffness of the superconducting state
and therefore its stability against fluctuations. 
\rhos also determines the superfluid density, a quantity that can be measured directly.

$\rho^{(s)}=\rho_s=(1/d)\tr\rhosm$ is easy to measure for 3D superconductors, given that it is related to 
to the London penetration depth $\lambda_L$ via the equation
\beq
 \lambda_L = \frac{\hbar}{e}\frac{1}{\sqrt{\mu_0\rho_s}}
\enq
where $\mu_0$ is the magnetic permeability.

For 2D superconductors $\rho_s$ cannot be obtained indirectly by measuring $\lambda_L$ and  recently techniques have been proposed to
obtain it via a direct measurement~\cite{Kapon2019}.
However, for 2D superconductors, and in general 2D ground states that break a U(1) symmetry,
$\rho^{\rm (s)}$ can also be obtained experimentally via Eq.~\ceq{eq:tkt} relating \tkt to $\rho^{\rm (s)}$.
In particular, for 2D superconductors, $T_{KT}$ can be obtained as the temperature at which 
the voltage $V$ across the superconductor scales as $I^3$, $I$ being the current.
This was the approach used in Ref.~\cite{Lu2019b} to estimate \tkt in MATBLG.
Using Eqs.~\ceq{eq:tkt} and \ceq{eq:ds_gen} we can relate \tkt to $\rho_s$. This requires to properly take into
account the temperature dependence of $\rho_s$: in addition to the temperature dependence due to the presence 
of the Fermi occupation factors, we must include the temperature dependence of the order parameter $\Delta$.
For many of the second-order phase transitions of interest, in first approximation, we can assume the "BCS scaling"
$\Delta(T)=1.76k_B(1-T/T_c)^{1/2}$.
In general $\Delta(T)$ can be obtained by solving the non-linear gap equation. For the concrete example of
superconducting MATBLG discussed in Sec.~\ref{sec:tblg} we have found that the BCS scaling of $\Delta(T)$ agrees well 
with the one obtained solving the non-linear gap equation.

\section{Quantum metric effects for correlated states in twisted bilayer graphene}
\label{sec:tblg}
The behavior of TBLG is particularly interesting for twist angles $\theta\sim 1.00\degree$.
For such small twist angles the moir\'e primitive cell is very large and the most effective
way to obtain the electronic structure is to use an effective low-energy continuum model~\cite{Bistritzer2011}.
The details of the model can be found in Ref.~\cite{Bistritzer2011}, here we briefly outline the model's essential elements and assumptions.
In graphene the conduction and valence bands cross at the corners ${\bf K}$ of the hexagonal Brillouin zone (BZ), 
$|{\bf K}|=4\pi/3a_0$ with $a_0$ the graphene's carbon-carbon distance.
Around the ${\bf K}$ points electrons in graphene behave as massless Dirac fermions~\cite{dassarma2011}
and the Hamiltonian for each layer, top (t) and bottom (b), forming TBLG is
\beq
 H_{t/b} = v_F \kk_{t/b}\cdot{\boldsymbol{\sigma}}-\mu\sigma_0,
 \label{eq:HamG}
\enq
where $v_F=10^6$~m/s is graphene's Fermi velocity, $\kk_{t/b}=(k_x, k_y)_{t/b}$ is the 2D momentum, measured from the ${\bf K}_{t/b}$ point,
for an electron in the top/bottom layer, ${\boldsymbol{\sigma}}=(\sigma_x,\sigma_y)$
is the 2D vector formed by the $x$, $y$ Pauli matrices in sublattice space~\cite{dassarma2011}, $\mu$ is the chemical potential, and $\sigma_0$
is the $2\times 2$ identity matrix.
Conservation of crystal momentum requires 
$\kk_b = \kk_t +(\KK_t-\KK_b) + (\GGv_t-\GGv_b)$. Here $\{\GGv_{t/b}\}$ are the reciprocal lattice
wave vectors in the top/bottom layer. Due to the twist the set of $\{\GGv_{t}\}$ is different from the set of $\{\GGv_{b}\}$.
In the model of Ref.~\cite{Bistritzer2011} only the tunneling processes for which 
$|\kk_b - \kk_t|=|\KK_t-\KK_b|=2K\sin(\theta/2)$, are taken into account.
There are three vectors ${\bf Q}_i = (\KK_t-\KK_b) +(\GGv_t-\GGv_b)_i$ ($i=-1,0,1)$ for which $Q\equiv|{\bf Q}|=2K\sin(\theta/2)$ 
to which correspond the interlayer tunneling matrices~\cite{Bistritzer2011}
\beq
 T_0 = w\begin{pmatrix} 1 & 1 \\ 1 & 1 \end{pmatrix}; \hspace{1cm} 
 T_{\pm 1} = w\begin{pmatrix} e^{\pm i2\pi/3} & 1 \\ e^{\mp i2\pi/3}  & e^{\pm i2\pi/3} \end{pmatrix}
\enq
where $w\approx 100$~meV is the interlayer tunneling strength. 
Up to an overall scale factor
the bands only depend on the ratio $w/v_F Q$~\cite{Bistritzer2011}.
In the remainder we set $w = 118$~meV.
The precise value of $w$ depends on the detail of the experimental sample.
In addition, due to corrugation effects the tunneling strength, $w_0$, for regions with AA stacking can be different from the one, $w_1$,
for regions with AB stacking. We assume the ratio $w_0/w_1$ to be uniform and equal to 1. 
Changes in the ratio $w_0/w_1$ affect the low energy bands and therefore
the superfluid stiffness.
All the tunneling processes for which $|\kk_b - \kk_t|=Q$ are taken into account by keeping all the recursive tunneling processes on a honeycomb structure
constructed in momentum space with nearest neighbor sites connected by the vectors ${\bf Q}_i$.
The primitive cell of this structure is the moir\'e lattice's mini-BZ. We adopt the 
convention in which the corners, $\kappa_{\pm}$, of the mini-BZ coincide with the points for which $\kk_{t/b}=0$. 
The number of sites of the honeycomb structure in momentum space used to obtain the band structure is increased until the bands converge. 
We find that for $w=118$~meV and $\theta\approx 1.00\degree$ convergence is reached when the number of sites is $\sim 200$.

\begin{figure}[htb]
 \begin{center}
  \centering 
  \includegraphics[width=\columnwidth]{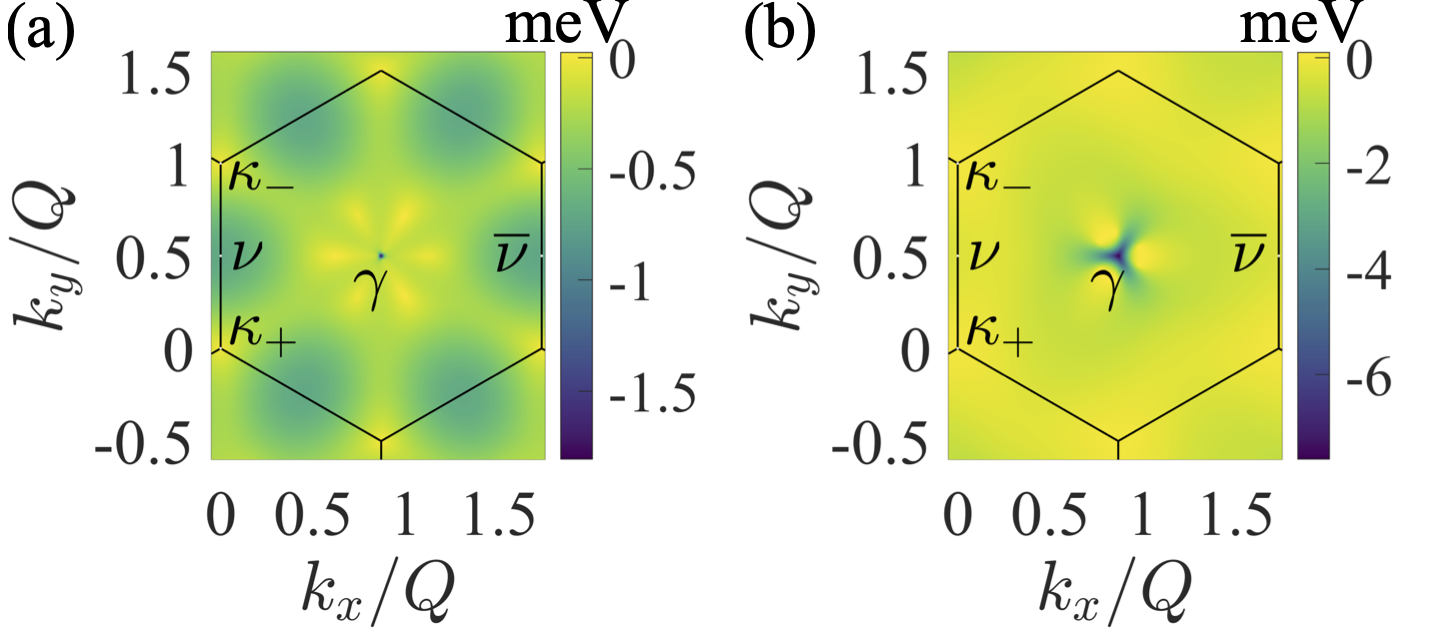}
  \caption{
  	Valence band of TBLG for $\theta=1.05\degree$, (a), and $\theta=1.00\degree$, (b). 
	The high symmetry points in the moir$\acute{\text{e}}$    Brillouin zone (BZ) are also shown.
	Adapted from~\cite{Hu2019}.
         } 
  \label{fig:bands}
 \end{center}
\end{figure}

In Ref.~\cite{Bistritzer2011} and other works $\theta_M$ is defined as the twist
angle for which the Fermi velocity at the $\kappa_{\pm}$ points of the mini-BZ vanishes, whereas in other works
it is defined as the value of $\theta$ for which the bandwidth of the conduction, or valence, band is minimum.
In the reminder we will adopt this second definition.
Figure~\ref{fig:bands}~(a) shows the 2D valence band at the magic angle $\theta=1.05\degree$.
We see that the bandwidth is just $\sim 2$~meV.
Small deviations of $\theta$ away from $\theta_M$
have large effects on the bandwidth of the lowest energy bands.
This can be seen from Fig.~\ref{fig:bands}~(b), showing the 2D valence band for $\theta=1.00\degree$:
a change of just $0.05\degree$ in $\theta$ results in a 
factor of 3 change in the bandwidth of the lowest energy bands.
The change in the bandwidth, in turn, strongly affects the stability, and properties of the correlated ground states.

The superconducting paring matrix $\hat\Delta$ is obtained via the mean-field approximation 
after adding an effective local (s-wave) attractive interaction whose strength is set so that 
at the magic angle, $\theta=1.05\degree$, $T_c=1.63$K when $\mu=-0.3$~meV~\cite{Hu2019}, in agreement with experiments~\cite{Cao2018}.
$\hat\Delta$ describes an s-wave superconductor whose only significant Fourier components
are the one with wave vector $\qq$ equal to zero and the ones with $\qq={\bf Q}_i$~\cite{Wu2018,Hu2019}.

\begin{figure}[htb]
 \begin{center}
  \centering 
  \includegraphics[width=\columnwidth]{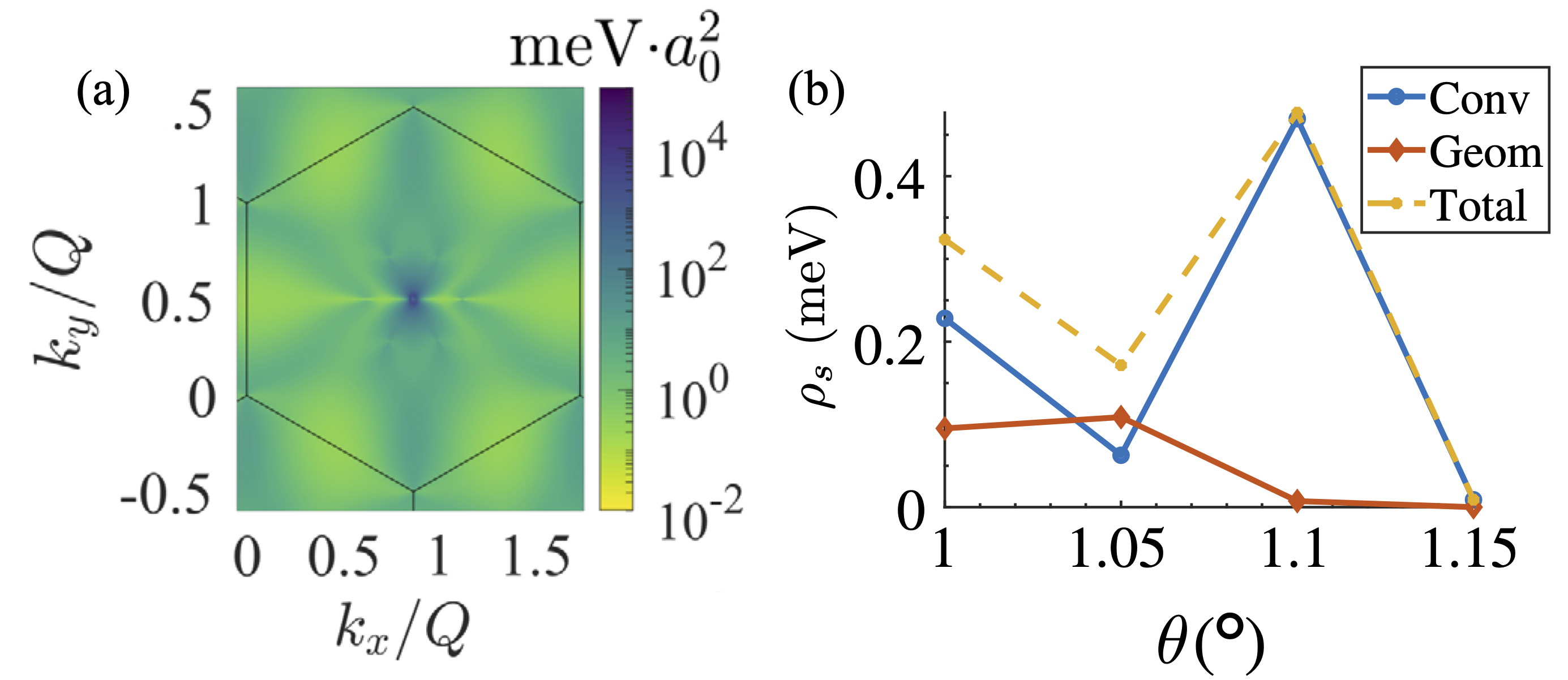}
  \caption{
  	       (a) Integrand of \rhoslg for TBLG at the magic angle. $\mu=-0.30$~meV. 
	       (b) Conventional (Conv) and geometric (Geom) contributions to the total longitudinal superfluid stiffness, $\rho_s\equiv (1/2)\tr\rhosm$, for TBLG
	       	   as a function of twist angle. $\mu=-0.3$~meV.
		   Adapted from~\cite{Hu2019}.
         } 
  \label{fig:ds_geo}
 \end{center}
\end{figure}

The large size of the moir\'e  primitive cell in TBLG when $\theta$ is of the order of $1\degree$
implies that effectively TBLG is a system with a large number of orbital degrees of freedom. 
This results in a very non-trivial quantum geometric tensors. In particular, for $\theta$
close to the magic angle, we have several regions of the BZ where the 
Berry curvature is very large. 
Considering that the positive semidefinite nature of \GG implies $\det g_{\mu\nu}\geq |B_{\mu\nu}|^2$, 
see Sec.~\ref{sec:general},
we expect
in these regions the geometric contribution
to \rhos to be large.  Fig.~\ref{fig:ds_geo}~(a) shows the profile in the BZ of the integrand
to obtain \rhoslg for $\theta=1.05\degree$.
From this figure we see that at the magic angle there are large regions in the mini BZ that provide
strong contributions to \rhoslg.
Figure~\ref{fig:ds_geo}~(b) shows the conventional, geometric, and total, longitudinal superfluid stiffness 
for different, small, values of $\theta$ and fixed $\mu$. 
We see that that \rhoslg is larger than \rhoslconv only close to the magic angle, but that it is significant
for all the values of $\theta$ smaller than $1.1\degree$.

\begin{figure}[htb]
 \begin{center}
  \centering 
  \includegraphics[width=\columnwidth]{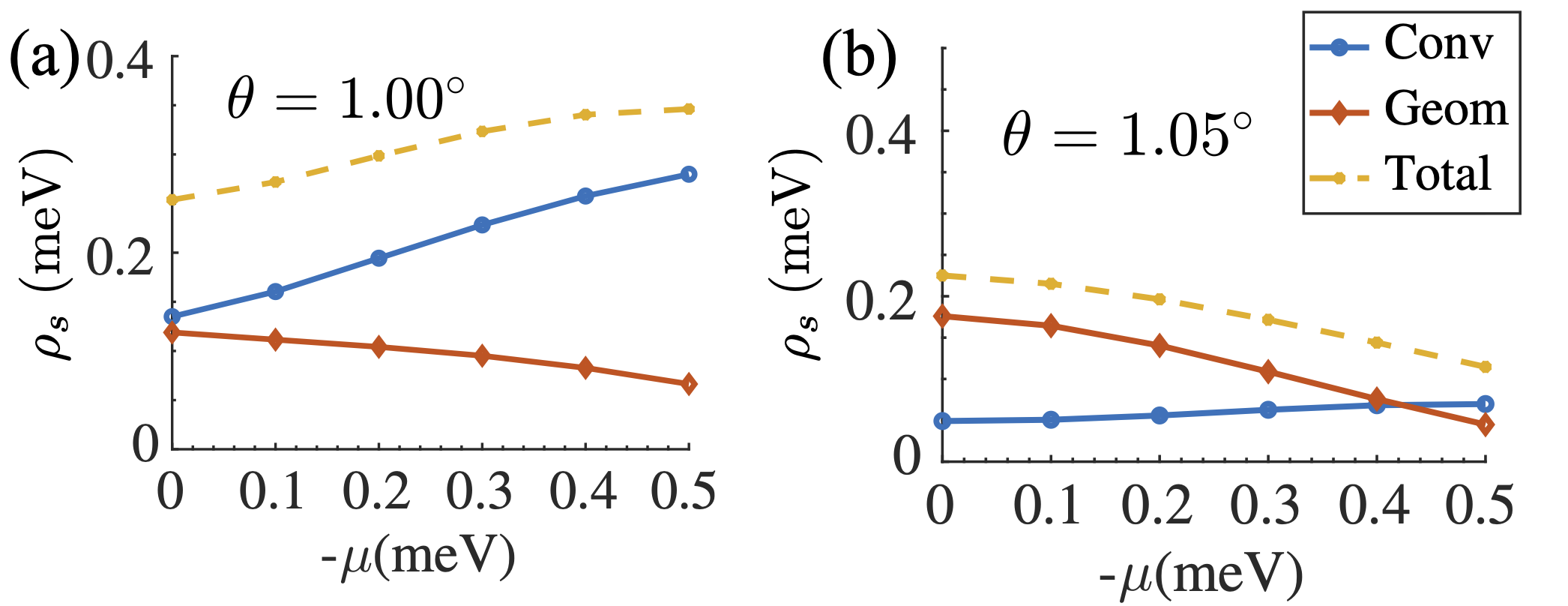}
  \caption{Conventional (Conv) and geometric (Geom) contributions to $\rho_s$ as a function of doping, $\mu$ (hole doping),
  		   for TBLG with $\theta=1.00\degree$, (a), and $\theta=1.05\degree$ (magic angle), (b).
  			Adapted from~\cite{Hu2019}.
         } 
  \label{fig:ds_vs_mu}
 \end{center}
\end{figure}

The results of Fig.~\ref{fig:ds_geo}~(b) show that systems like TBLG are an ideal playground in which to test
the connection between quantum geometry and macroscopic properties of correlated ground states.
This can be seen, for instance, by considering the scaling of $\rho_s$ with the chemical potential $\mu$
at the magic angle, and away from it.
The conventional contribution to \rhos, in general, increases with doping, and therefore with $\mu$.
As a consequence, in systems in which the superfluid stiffness is mostly due to the conventional term,
the total $\rho_s$ increases with $\mu$. This is the case also for TBLG away from the magic angle
as shown in Fig~\ref{fig:ds_vs_mu}~(a) for which
the conventional contribution to $\rho_s$ is larger than the geometric contribution.
The geometric contribution to \rhos, in general, can increase or decrease with doping.
From Fig.~\ref{fig:ds_vs_mu}~(a) we see that, for $\theta=1.00\degree$, \rhosg
decreases with $\mu$. 
This is also the case at the magic angle where, however, \rhosg dominates over \rhosconv.
As a consequence at the magic angle we have the unusual situation
that the total \rhos decreases with $\mu$ as
as shown in Fig~\ref{fig:ds_vs_mu}~(b). 

\begin{figure}[htb]
 \begin{center}
  \centering 
  \includegraphics[width=\columnwidth]{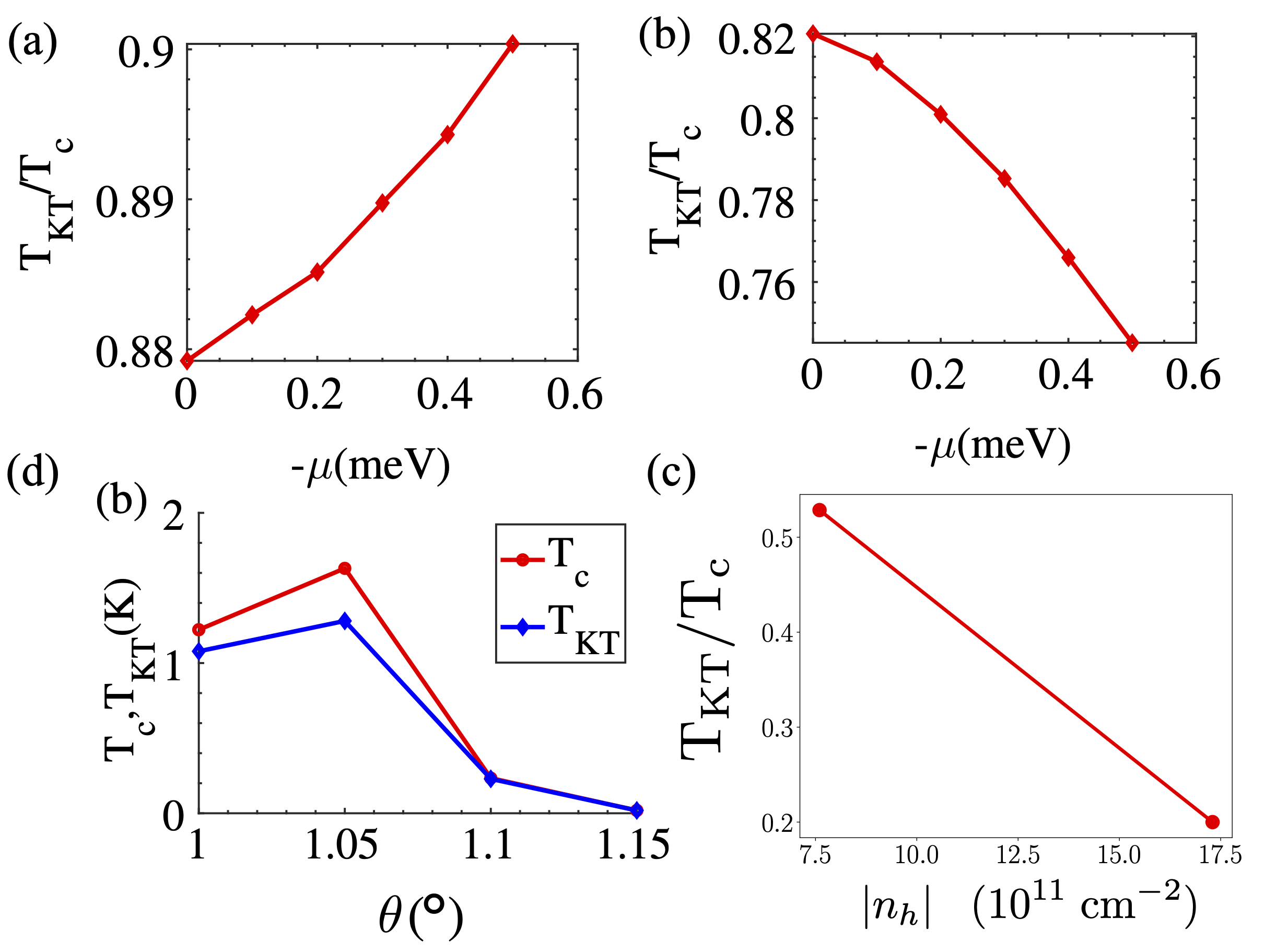}
  \caption{
  	      Calculated $T_{KT}/T_c$ as a function of $\mu$ for $\theta=1.00\degree$, away from the magic angle, (a), and 
	      at the magic angle (b);  adapted from~\cite{Hu2019}.
	      (c) $T_{KT}/T_c$ as a function of $|n_h|$ in the hole-doped regime obtained from the experimental	
	      measurements presented in Ref.~\cite{Lu2019b}.
	      (d) $T_c$ and $T_{KT}$ as a function of $\theta$ for TBLG when $\mu=-0.3$~meV; adapted from~\cite{Hu2019}.	          
         } 
  \label{fig:tkt}
 \end{center}
\end{figure}

We expect that the scalings of $\rho_s$ with respect to $\mu$ will be reflected in the scaling of \tkt.
Using Eq.~\ceq{eq:tkt}, knowing the temperature scaling of \rhos, \tkt can be calculated.
Figure~\ref{fig:tkt}~(a) shows the results for the ratio \tkt$/T_c$ away from the magic angle, $\theta=1.00\degree$.
As expected we see that \tkt$/T_c$ increases as the hole density increases.
At the magic angle we have instead that \tkt$/T_c$ decreases with doping, as shown in Fig~\ref{fig:tkt}~(b),
a consequence of the fact that at the magic angle the geometric contribution of \rhos dominates.

In a 2D superconductor the unbounding of the vortices due to thermal fluctuations causes a finite resistance
and therefore a finite longitudinal voltage, $V_{xx}$, that depends on the strength of the electrical current $I$ driven through the system.
For $T=T_{KT}$ we have that $V_{xx}\propto I^3$. By measuring the $V_{xx}(I)$ relation at different temperatures
is then possible to estimate \tkt as the temperature for which $V_{xx}\propto I^3$. For TBLG this was
done in Ref.~\cite{Lu2019b}. 
Figure~\ref{fig:tkt}~(c) shows the scaling of \tkt$/T_c$ obtained using the two data points presented 
in the "Extended Data Table 1"  of Ref.~\cite{Lu2019b} for MATBLG in the hole doped regime.
The figure shows that in the MATBLG samples used in Ref.~\cite{Lu2019b}, in the hole-doped regime,  \tkt$/T_c$
decreases with doping in qualitative agreement with the results of Fig~\ref{fig:tkt}~(b),
suggesting that also experimentally, in the hole-doped regime, the geometric contribution of \rhos dominates
over the conventional one.

Figure~\ref{fig:tkt}~(d) shows the dependence in TBLG of \tkt on the twist angle for fixed chemical potential, $\mu=-0.3$~meV.
These results show that, because of the geometric contribution to $\rho_s$, at the magic angle 
\tkt is largest, along with $T_c$. This suggests that in multiorbital systems, like TBLG, the geometric contribution to $\rho_s$
can compensate the suppression of \rhoslconv associated with the flattening of the bands and lead to robust superfluid states.

The discussion above focused on the case when the correlated ground state breaking a $U(1)$ symmetry is the superconducting state.
A very similar discussion can be carried out for other ground states that break a $U(1)$ symmetry. 
In particular similar results can be obtained for the ferromagnetic state~\cite{Alavirad2020,Wu2020c,Bernevig2020a}.
Recently it has been suggested that an "orbital-magnetic" state, characterized by a non-zero sublattice polarization, 
might be one of the correlated states most likely realized in 
TBLG~\cite{Sharpe2019,Xie2020,Serlin2020a,Wu2021a}. 
Also for this state,
an analysis similar to the one presented above for the superconducting state can be done.

More recently, we have considered the possibility that in double TBLG an exciton condensate state might be realized~\cite{Hu2020}. 
This is a long sought correlated
state in which electron and holes (e-h) pair to form a neutral 
superfluid~\cite{Keldysh1965,Halperin1968,lozovik1975,lozovik1976,eisenstein2004,Fogler2014a,Gupta2020,Wang2019b,Wang2021}.
We considered a double layer formed by two MATBLG, one electron-doped and one hole-doped,
separated by a thin dielectric.
As  for the case of superconductivity the flatness of the bands, while favoring
the formation of e-h pairs, can lead to a very small superfluid density. 
We found that, for the exciton condensate, the quantum metric plays an even more critical  
role than for the superconducting case in stabilizing the collective state and in
guaranteeing a nonzero value of the superfluid stiffness~\cite{Hu2020}.

\section{Conclusions and Outlook}
\label{sec:outlook}
The experimental realization of magic angle twisted bilayer graphene systems has opened a completely new avenue to explore
the connection between the metric of quantum states and the properties of strongly correlated states that break
continuous symmetries. It has shown experimentally that the flatness of the low energy bands does not necessarily imply
a low superconducting density \rhos and demonstrated the importance of the interband contributions, associated with
a non-trivial quantum metric of the bands, to \rhos.  

The experimental results on MATBLG, combined with the theoretical treatment of \rhos that includes the geometric 
contribution~\cite{Peotta2015,Julku2016,Liang2017,Hazra2019,Hu2019,Xie2019,Julku2020,Wu2020,Bernevig2020a},
show that the quantum metric plays an important role in determining the properties of the correlated states of multi-orbital systems.
Multilayers formed by 2D crystals stacked with relative small twist angles have very large moir\'e primitive cells and therefore
many orbitals and low energy bands with very small bandwidths. 
For these systems, therefore, the quantum metric plays in important role in determining the stability
and properties of correlated ground states. 
We expect that the study of the connection between quantum metric and properties of correlated states will
be extended to several new twisted 2D multilayers, both based on graphene~\cite{Chen2019,Park2021}, and on
other 2D crystals such as monolayers of transition metal 
dichalcogenides~\cite{abergel2013,jzhang2013,jzhang2014,Triola2016,Liu2014,Wu2019a,rodriguez2019,Gani2019,Wang2020,Regan2020,Zhang2020,Xu2020z,Rossi2019a,Ghiotto2021}.

A new interesting research direction would be the study of 
the interplay between quantum metric, disorder, and stiffness of the correlated states,
in particular in twisted bilayers~\cite{lu2016}.
We could expect that for states like superconductivity disorder might suppress 
the conventional part of \rhos more than the geometric part. It will be interesting to verify
theoretically and experimentally the extent of the validity of such expectation.

Correlated states that break a continuous symmetry can differ topologically.
For these states it will be interesting to investigate how the connection between quantum metric and stiffness
might vary between the different topological phases, and, more in particular, 
if there are features of such connection that can be used to identify the 
topological phases. For instance, topologically different superconducting phases
can be realized in superconducting quantum anomalous Hall (QAH) states~\cite{Qi2010a}.
Considering the recent observation of signatures of QAH states in MATBLG, superconducting topological states
might be realized in MATBLG proximitized to a superconductor. 

In some cases, correlated states breaking different continuous symmetries can compete or coexist. 
It will be interesting to study the relation between quantum metric and properties such as \rhos
of competing or coexisting collective states in systems like TBLG. 

As discussed in Sec.~\ref{sec:superfluid}, in 2D systems, experimental evidence of the connection between
quantum metric and \rhos can be obtained indirectly by obtaining the scaling of \tkt with 
respect to other tunable quantities such as doping. It will be interesting
to have more direct experimental evidence of the effects of the metric of the quantum
states on the properties of correlated states. 
One approach would be to measure the dispersion of the Goldstone modes
associated with the spontaneous breaking of the continuous symmetry given that \rhos
enters the dispersion of such modes.

In general, the quantitative understanding of the relation between quantum metric and the stability
and properties of collective ground states will allow to better design strongly interacting systems
with the desired functionalities. By designing multiorbital systems with flat bands
that maximize the quantum metric we can achieve both large values of $T_c$ and superfluid density,
properties that are desirable in several applications.

\section{Acknowldgments}
It is a pleasure to thank the collaborators with whom I have had the opportunity to collaborate 
in the past few years on topics related to the subject discussed in this article: 
Yafis Barlas, Xiang Hu, Timo Hyart, Alexander Lau, Sebastiano Peotta, and Dmitry Pikulin.
The work has been supported by NSF CAREER Grant No. DMR-1455233, and ARO Grant No. W911NF-18-1-0290.
The author also thanks KITP, supported by Grant No. NSF PHY1748958, where part of this work was performed.
%


%


\end{document}